\documentclass{parcfd}

\usepackage{nopageno}
\usepackage{graphicx}
\usepackage{amsmath}
\usepackage{amsfonts}
\usepackage{amssymb}
\usepackage{hyperref}
\usepackage{layout}
\usepackage{color}
\usepackage{lmodern}  
\usepackage[T1]{fontenc}

\usepackage{booktabs} 
\usepackage{subcaption}

\title{SCALING WATERLILY.JL WITH MPI AND AN IMPROVED GEOMETRIC MULTIGRID SOLVER}

\author{Bernat FONT$^{*}$, Marin LAUBER, Tzu-Yao HUANG, \\Gabriel D. WEYMOUTH}

\heading{Bernat Font, Marin Lauber, Tzu-Yao Huang, Gabriel D. Weymouth}

\address{Delft University of Technology\\
Faculty of Mechanical Engineering\\
Mekelweg 2, 2628 CD Delft, The Netherlands\\
$^{*}$ e-mail: \texttt{b.font@tudelft.nl}, web page: \texttt{https://b-fg.github.io}}


\abstract{
We present recent performance-oriented developments in WaterLily.jl, a scale-resolving incompressible flow solver written in pure Julia that runs seamlessly on CPUs and GPUs of any vendor.
Supported by the newly added MPI-based parallelism, strong-scalability tests display a near-ideal linear trend, and weak-scaling efficiency is kept above 85\% before node memory-concurrency contention dominates parallel performance.
Inter-node weak scalability is sustained above 96\% with grid size up to 1 billion cells.
We further benchmark improvements to the geometric multigrid Poisson solver enabled by an adaptive under-relaxed red-black Gauss--Seidel smoother together with anisotropic coarsening operators.
}

\begin{document}


\vspace{-0.6cm}
\section{INTRODUCTION}

High-performance computing (HPC) is a cornerstone of computational fluid dynamics, enabling an ever-growing resolution in scale-resolving simulations.
Driven by hardware advances such as general-purpose processor (CPU) and accelerator-based (GPU) nodes, CFD software is rapidly evolving to harvest the increase in computational power.
The trend comes at an added software-complexity cost, and CFD developers now heavily rely on HPC libraries to offload compute-expensive kernels and parallelisation.

In this direction, we present performance-oriented developments and the first parallel scalability results of \texttt{WaterLily.jl} \cite{WeymouthFont2025}.
Written in pure Julia, WaterLily is designed to be a compact and minimalist finite-volume incompressible-flow solver that runs on any CPU or GPU computing backend.
This feature is enabled by separating computing workloads (typically \texttt{for} loops) from the algorithmic space-time discretization.
By using Julia's metaprogramming support, the solver ultimately specialises each kernel to the available computing backend using \texttt{KernelAbstractions.jl}\cite{Churavy_KernelAbstractions_jl}.

\section{MPI PARALLELISATION AND SCALING}

Further leveraging Julia's scientific computing ecosystem, distributed-memory parallelism has been recently integrated in WaterLily via the \texttt{ImplicitGlobalGrid.jl} (IGG) package \cite{Omlin2024}.
Backed by the \texttt{MPI.jl} wrapper, IGG uses a Cartesian MPI communicator naturally mapping to regular grids, and offering CUDA-aware or ROCm-aware MPI support for GPU-based applications.
Additionally, IGG works with different grid sizes, thus enabling halo updates within geometric-multigrid (GMG) levels without added complexity from the IGG-user side.

\begin{figure}[!t]
\centering
\includegraphics[width=7.93cm]{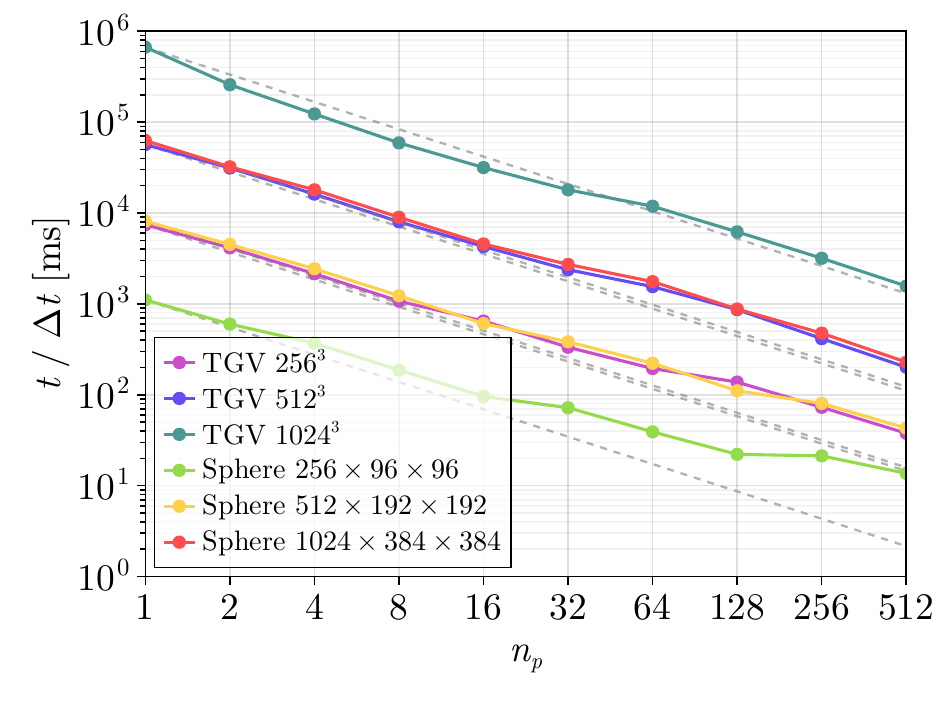}
\includegraphics[width=7.93cm]{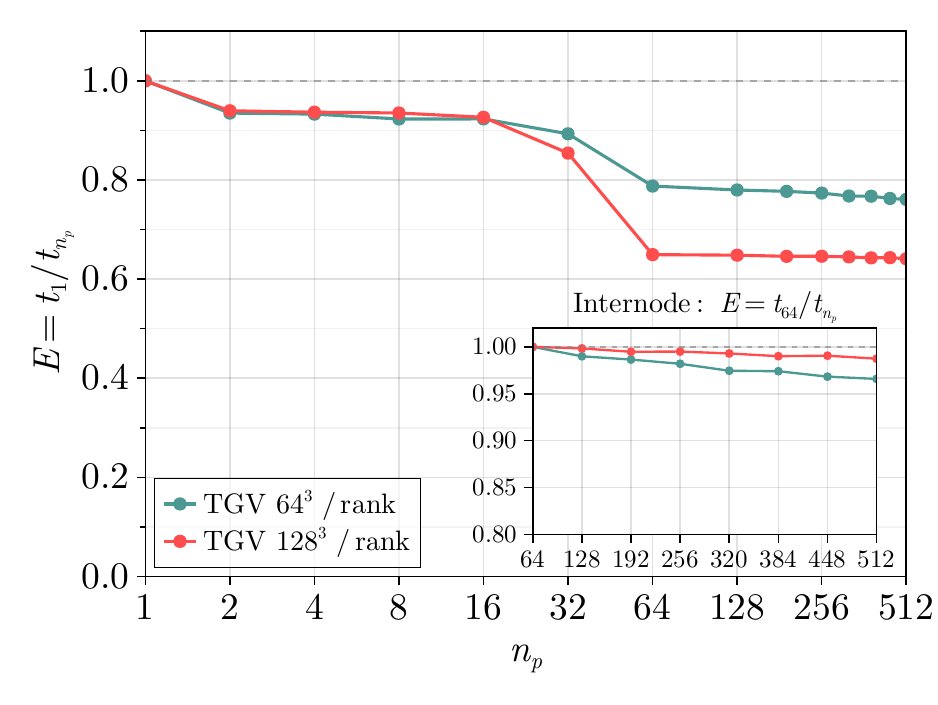}
\caption{Strong scaling (left) and weak scaling (right) results on the Rome CPU partition of the Snellius Dutch national supercomputer. The inset shows the inter-node weak-scaling efficiency, normalised by the single-node (64-rank) time.}
\vspace{-0.5cm}
\label{fig:scalability}
\end{figure}

Strong and weak scalability results are presented in Figure~\ref{fig:scalability}.
The tests are conducted with the SIMD single-thread CPU backend with single precision (FP32) and half-node occupancy (64 of the 128 cores of the dual-socket AMD EPYC 7H12 nodes).
The test cases comprise a Taylor--Green vortex (TGV) at $Re=1600$ in a triple-periodic cubic domain, and flow past a sphere at $Re_D=3700$ in a $(16\times6\times6)D$ domain.
The MPI domain decomposition always minimizes the halo surface for a given number of processes.
The strong-scalability results closely follow the ideal linear reduction in time per time-step, even super-linearly for the largest grids as shrinking per-rank load relaxes the memory-concurrency contention discussed below.
The trend deteriorates when MPI halo updates dominate the local workload, once the smallest local subdomain dimension approaches ${\sim}16$ cells. The weak-scalability results maintain an efficiency $(E)$ above 90\% for the small ($64^3$ cells) and large ($128^3$ cells) loads up to 16 ranks.
Beyond, efficiency drops due to intra-node memory-concurrency contention, motivating the 50\% node occupancy.
This reflects the memory-bound nature of the solver's stencil kernels \cite{Williams2009}.
While memory bandwidth is not saturated (checked with hardware counters), ranks compete to access the per-chiplet link to the memory controllers.
Ranks are hence pinned with a uniform stride across the node.
Beyond the node, weak scaling is nearly ideal, remaining above 96\% up to 512 ranks and $10^9$ cells (Figure~\ref{fig:scalability}, inset).

\section{ANISOTROPIC GEOMETRIC MULTIGRID SOLVER WITH ADAPTIVE UNDER-RELAXED RED-BLACK GAUSS--SEIDEL SMOOTHER}

WaterLily's general-purpose Poisson solver is Preconditioned Conjugate Gradient (PCG), which applies to arbitrary grid sizes and extends naturally to parallel backends. However, its global inner products require costly reductions in MPI simulations, and the number of iterations increases with problem size, resulting in poor weak scaling. For this reason, the default WaterLily solver uses GMG, which accelerates convergence by combining smoothers to remove high-frequency errors with coarse-grid correction (CGC) for low-frequency errors. The GMG solver has been further optimized with the development of an adaptive smoother and anisotropic coarsening.

\subsection{Adaptive under-relaxed Red-Black Gauss--Seidel}

While PCG can be used as a smoother, its non-local behaviour conflicts with CGC, particularly for the variable-coefficient Poisson equations arising in multiphase and immersed-boundary flows. In contrast, Red-Black Gauss--Seidel (RBGS) naturally damps high-frequency errors and avoids the global reductions required by PCG, making it well suited to parallel multigrid. To ensure monotonic residual decay in highly heterogeneous media, we introduce a dynamically under-relaxed RBGS smoother whose relaxation factor adapts to residual growth or decay after each V-cycle. Numerical tests (Table~\ref{tab:side_by_side_comparison}) show that this adaptive GMG-RBGS solver maintains robust convergence while reducing solver cost by 1.6–2.3× compared with a PCG smoother, and by 3.7–5.8× compared with the generic PCG solver.

\begin{table}[!t]
\centering
\caption{Solver comparison for the case of a simplified swimming jellyfish. Grid size is $4\times2^{3p}$. Cost reported in $\mathrm{ns}$ per cell per time-step running on a laptop RTX 4060 GPU.}
\label{tab:side_by_side_comparison}
\footnotesize 
\setlength{\tabcolsep}{4pt} 
\begin{subtable}[t]{0.48\textwidth}
\centering
\begin{tabular}{llccc}
\toprule
$p$ & \textbf{Solver} & \textbf{\# Iter.} & \textbf{Final resid.} & \textbf{Cost} \\
\midrule
5 & PCG & 14.90 & $9.45 \times 10^{-5}$ & 139.84 \\
 & GMG-PCG & 1.75 & $3.09 \times 10^{-5}$ & 86.97 \\
 & GMG-RBGS & 1.70 & $5.44 \times 10^{-5}$ & 38.22 \\
\bottomrule
\end{tabular}
\end{subtable}
\hfill
\begin{subtable}[t]{0.48\textwidth}
\centering
\begin{tabular}{llccc}
\toprule
$p$ & \textbf{Solver} & \textbf{\# Iter.} & \textbf{Final resid.} & \textbf{Cost} \\
\midrule
6 & PCG & 26.09 & $9.43 \times 10^{-5}$ & 95.93 \\
 & GMG-PCG & 1.89 & $5.33 \times 10^{-5}$ & 26.25 \\
 & GMG-RBGS & 2.34 & $5.69 \times 10^{-5}$ & 16.61 \\
\bottomrule
\end{tabular}
\end{subtable}
\end{table}

\subsection{Anisotropic multigrid coarsening}

\begin{figure}[!ht]
    \centering
    \includegraphics[width=1\linewidth]{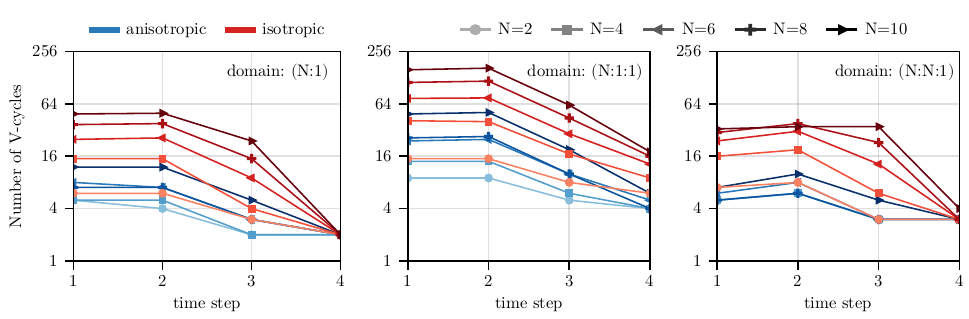}
    \caption{Total number of V-cycle (predictor + corrector steps) for the first 4 time steps of canonical impulsive-flow problems: (left) 2D flow around a circular cylinder, (center) 3D flow around a spanwise-periodic cylinder and (right) 3D flow around a sphere. The different domain aspect ratios are given as a small insert in the top-right corner of each subplot.}
    \label{fig:semicoarsening}
\end{figure}

With isotropic coarsening, the smallest domain dimension limits the number of multigrid levels, bounding the decay of low-frequency error.
For highly anisotropic domains, this can severely penalize the convergence of the Poisson solver.
This is especially important for impulsive flow cases, where the low wavenumber axial pressure wave remains at an intermediate wavenumber on the coarsest level, hindering the smoother's efficacy.
Instead, anisotropic coarsening downsamples every dimension independently and reaches the coarsest domain size where the low wavenumber content is easily removed by the smoother.
Implemented in WaterLily, the anisotropic multigrid coarsening leverages the Cartesian structure of the grid with minimal overhead compared to the classical isotropic approach.
In contrast to isotropic coarsening, we find that this approach allows us to effectively reduce the number of multigrid cycles for impulsive flow problems on high aspect-ratio rectangular grids, as displayed in Figure~\ref{fig:semicoarsening}.
For domains with aspect ratio $N\ge4$, approximately an order of magnitude fewer V-cycles are required to converge to the same residual.
This trend is maintained for all cases and aspect ratios tested.

\section{CONCLUSIONS}

Recent developments in \texttt{WaterLily.jl} have improved the Poisson solver through aniso-\\tropic GMG coarsening and an RBGS smoother, targeting both faster convergence and total solve wall-time reduction.
Together with the newly implemented distributed-memory parallelism, tested in strong and weak-scalability setups, WaterLily is now ready to leverage many-GPU architectures and embrace the exascale computing challenge.
This opens new research avenues in scale-resolving simulations including vortex-dominated flows, fluid-structure interaction problems, and optimization.

\bibliographystyle{ieeetr}
\bibliography{main}

\end{document}